\documentclass[aps,prb,reprint,showpacs,superscriptaddress]{revtex4-1}

\usepackage{amsmath}
\usepackage{epsfig}
\usepackage{epstopdf}
\usepackage{textcomp}

\begin{document}

\title{Spectroscopic mapping of local structural distortions in ferroelectric PbTiO$_3$/SrTiO$_3$ superlattices at the unit-cell scale}

Published in Physical Review B vol \textbf{84}, 220102 (2011)
\author{Almudena~Torres-Pardo}
\email{almudena.torres@u-psud.fr, atorresp@quim.ucm.es}
\affiliation{Laboratoire de Physique des Solides, Universit\'e Paris-Sud, CNRS-UMR 8502, Orsay 91405, France}
\affiliation{DPMC, University of Geneva, 24 quai Ernest-Ansermet,
1211 Geneva 4, Switzerland}
\author{Alexandre~Gloter}
\email{alexandre.gloter@u-psud.fr}
\affiliation{Laboratoire de Physique des Solides, Universit\'e Paris-Sud, CNRS-UMR 8502, Orsay 91405, France}
\author{Pavlo~Zubko}
\author{No\'emie~Jecklin}
\author{C\'eline~Lichtensteiger}
\affiliation{DPMC, University of Geneva, 24 quai Ernest-Ansermet,
1211 Geneva 4, Switzerland}
\author{Christian~Colliex}
\affiliation{Laboratoire de Physique des Solides, Universit\'e Paris-Sud, CNRS-UMR 8502, Orsay 91405, France}
\author{Jean-Marc~Triscone}
\affiliation{DPMC, University of Geneva, 24 quai Ernest-Ansermet,
1211 Geneva 4, Switzerland}
\author{Odile~St\'ephan}
\affiliation{Laboratoire de Physique des Solides, Universit\'e Paris-Sud, CNRS-UMR 8502, Orsay 91405, France}

\date{\today}

\begin{abstract}
The local structural distortions in polydomain ferroelectric PbTiO$_{3}$/SrTiO$_{3}$ superlattices are investigated by means of high spatial and energy resolution electron energy loss spectroscopy combined with high angle annular dark field imaging. Local structural variations across the interfaces have been identified with unit cell resolution through the analysis of the energy loss near edge structure of the Ti-L$_{2,3}$ and O-K edges. \textit{Ab-initio} and multiplet calculations of the Ti-L$_{2,3}$ edges provide unambiguous evidence for an inhomogeneous polarization profile associated with the observed structural distortions across the superlattice.
\end{abstract}
\maketitle
Complex oxide heterostructures offer a vast playground for exploring and combining the many functional properties of these fascinating materials arising from the subtle interplay between their charge, spin, orbital and lattice degrees of freedom. Bilayers, multilayers and superlattices composed of ultrathin oxide layers not only shed new light on our fundamental understanding of the constituent materials, but frequently reveal unexpected new phases at their interfaces \cite{ZubkoARCMP2011}. Superlattices composed of ferroelectric and paraelectric oxides have been the subject of numerous studies, motivated by fundamental questions about ferroelectric size effects, by the possibilities these artificially layered materials offer for tailoring their functional properties, and by the fascinating interface physics they display. Ultrafine period superlattices, composed of ferroelectric PbTiO$_3$ (PTO) and paraelectric SrTiO$_3$ (STO), for example, have been shown to exhibit improper ferroelectricity driven by the coupling of the polar and non-polar distortions at the interface \cite{Bousquet2008}. More recently, regular ferroelectric nanodomains have been observed in PTO/STO  superlattices with larger periodicities and were shown to be responsible for large enhancements in the effective dielectric constant \cite{ZubkoPRL2010}. Such domains are expected to give rise to complex inhomogeneous structural distortions and polarization profiles \cite{LiAPL2007,LisenkovPRB2007}, departing from uniform polarization models, frequently used to describe the properties of ferroelectric/paraelectric superlattices in the absence of domains \cite{NeatonRabe2003}.\\
\- A microscopic insight into the local structure is thus key to understanding the behavior of these exiting artificially layered materials, and here transmission electron microscopy (TEM), with the recent advances in spatial resolution, provides an invaluable tool. Individual ionic displacements within a single perovskite unit cell can now be identified from phase contrast images. The spatial resolution is high enough to determine the direction and even the magnitude of the local dipole moments in ferroelectric materials \cite{JiaNatMater2007,JiaNature2008,ChisholmPRL2010}, and has recently enabled the direct verification of the existence of polarization rotation at domain walls in Pb(Zr,Ti)O$_3$ ferroelectric thin films \cite{JiaScience2011}.\\
\begin{figure}
\includegraphics[width=\columnwidth]{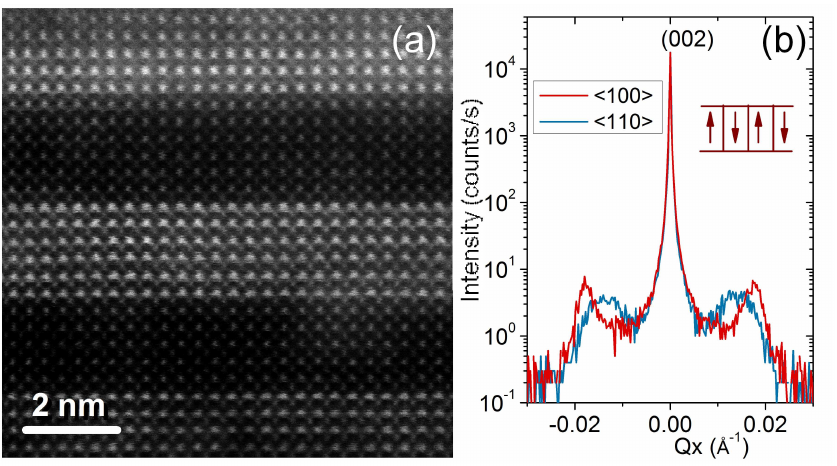}
\caption{(a) HAADF-STEM image of the [010] projection of a (6$|$6) PTO/STO superlattice (b) X-ray reciprocal space scans around the (002) superlattice reflection show broad satellite peaks due to 180$\,^{\circ}$ ferroelectric domains.}
\end{figure}
In this article, we focus on an alternative spectroscopic technique to study local ferroelectric distortions at the single unit-cell scale. Atomically-resolved electron energy loss spectra (EELS) and high angle annular dark field (HAADF) images have been obtained to access the local structure \cite{BruleyJMicros2001,ZhangPRB2005,HaturaPRB2009,CravenUltramicroscopy2010,BlelochJApplPhys2011} of polydomain PTO/STO superlattices. We show that the energy loss near edge fine structure (ELNES) of the Ti-L$_{2,3}$-edge is highly sensitive to very small atomic displacements ($<0.1$\,\AA). Changes in the tetragonal distortion of the perovskite unit cell as small as 1\% have been detected, revealing a variation of tetragonality within individual PTO layers and thus providing direct evidence of local structural inhomogeneities in polydomain ferroelectric/paraelectric superlattices. \textit{Ab-initio} and charge transfer multiplet calculations show distinct differences between the spectra expected for polar and non-polar structures, demonstrating that the EELS technique has sufficient energy and spatial resolution to probe ferroelectricity at the unit-cell scale.

We have studied a $(6|6)_{21}$ superlattice with 21 bilayers of 6 unit cells of PTO and 6 unit cells of STO on top of a single crystalline STO substrate. Epitaxial top and bottom SrRuO$_3$ electrodes were also deposited \textit{in situ}. Details of sample growth and electrical characterization can be found in Ref.~\onlinecite{ZubkoPRL2010}.
Atomically-resolved HAADF images and EELS spectra were acquired with an aberration-corrected scanning transmission electron microscope (STEM) Nion UltraSTEM 100 \cite{KrivanekUltra2008}. Figure 1a displays a HAADF image of the superlattice projected onto the (010) plane. The variation in HAADF signal intensity for each atomic column reflects the difference in atomic numbers between Pb (Z=82) and Sr (Z=38) cations \cite{PennycookUltram1989}, demonstrating the atomic sharpness of the interfaces. X-ray diffraction measurements reveal a periodic in-plane modulation attributed to 180\textdegree\ ferroelectric domains (Fig.1b). The domain satellites are observed independent of the in-plane orientation of the sample, indicating an almost isotropic distribution of domain wall alignments. The domain periodicity ranges from 55~\AA\ (for domain walls along $\langle 100\rangle$) to 65~\AA\ (along $\langle 110\rangle$). This means that in general, the domain wall orientation will be random with respect to the imaging plane and the HAADF and EELS images will average over the domains within the thickness of the TEM specimen.\\
\begin{figure}
\includegraphics[width=\columnwidth]{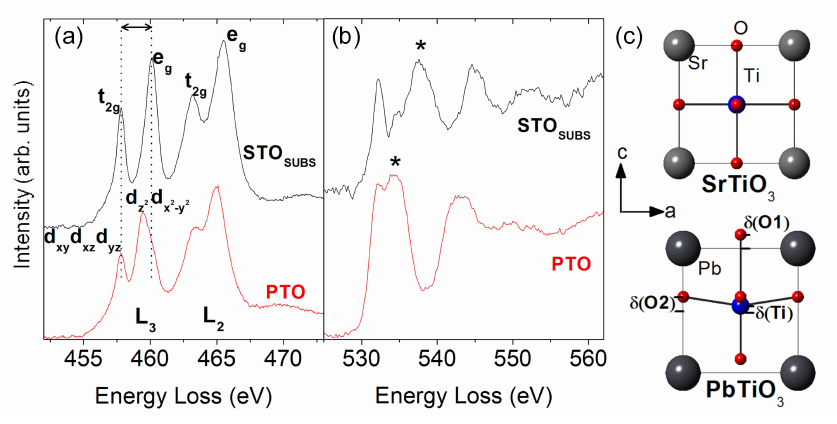}
\caption{(a) Ti-L$_{2,3}$ and (b) O-K edges for STO substrate (black-upper) and 12 nm thick PTO layer (red-lower). (c) Schematic projection along the [010] zone-axis of the STO and PTO unit cells. The displacement of the atomic position is displayed as $\delta${(Ti)}, $\delta${(O1)} and $\delta${(O2)}.}
\end{figure}
As a reference for the characterization of the $(6|6)$ superlattice, Ti-L$_{2,3}$ and O-K EELS spectra were acquired from a 12~nm thick PTO layer and from the STO substrate (7 nm away from the interface).
Figure 2a shows the Ti-L$_{2,3}$ ELNES of the reference spectra for PTO and STO to be compared with those of the superlattice.
The energy resolution (0.4~eV) achieved allows us to identify slight variations between the STO and PTO Ti-L$_{2,3}$ ELNES that can be correlated with structural differences between the two oxides.
While the unit cell of the STO substrate has cubic symmetry ($a$=$b$=$c$=3.905 \,\AA), the hybridization between the Pb 6\textit{s} and the O 2\textit{p} states induces a large tetragonal strain \cite{CohenNature1992} in the PTO unit cell ($c/a$=1.063) (Fig.2c).
Due to the reduction in symmetry from the cubic perovskite structure, the crystal field splitting (CFS) of the Ti 3\textit{d} e$_g$ and t$_{2g}$ orbitals is expected to decrease, leading to the observed reduction in the L$_3$-edge energy splitting ($\Delta$L$_3$) \cite{GrootPRB1990,GreedanJSolidState2005} from 2.3 eV in the STO substrate to 1.65 eV for the PTO layer (Fig.2a).
The site-symmetry of the Ti atom in the TiO$_{6}$ octahedra is also reduced in the PTO unit cell, the corresponding Ti L$_3$-e$_{g}$ peak showing an asymmetric broadening towards higher energies (see Fig.2a), that has been attributed to the non-centrosymmetric location of the Ti cation in the TiO$_{6}$ octahedra \cite{JanAPL2003,BlelochJApplPhys2011}.

Comparing the O-K ELNES of the STO substrate and the 12~nm thick PTO layer in Fig.2b, clear differences between both spectra can be identified. The most pronounced one is the large shift observed in the
position of the second peak (marked with an asterisk). This peak has been attributed to the hybridization of the O 2\textit{p} with Sr 4\textit{d} and with Pb 6\textit{sp} states for STO and PTO, respectively \cite{FuAPL2005,FittingUltramicroscopy2006,MullerPhiMag2010}, making it possible to distinguish between Sr and Pb containing cells by identifying the position of this peak. The analysis of the Ti-L$_{2,3}$ and O-K edges therefore offers an excellent method for determining both structural and chemical variations within the PTO/STO superlattices.

Figure 3a shows a series of 512 individual EELS spectra acquired in spectrum-line mode with dwell time=125 ms, displaying the Ti-L$_{2,3}$ and the O-K edges across four interfaces of the (6$|$6) superlattice. The high spatial resolution of STEM allows us to identify the core loss edges from individual atomic columns, while the dark field profile is simultaneously recorded along the scanning line (black line in Fig.3b).
The Ti-L$_{2,3}$ and O-K EELS spectra corresponding to 7 consecutive unit cells across one of the PTO-STO interfaces (numbered from 6 to 12 in the image) are displayed in Fig.3d and 3e, respectively. Each spectrum is the sum of 22 individual spectra over one unit cell.
In order to improve the energy resolution, the Ti-L$_{2,3}$ spectra in Fig.3f have been deconvoluted using the Richardson-Lucy (R-L) approach \cite{GloterUltramicr2003}.

Consistent with their large tetragonality, the Ti-L$_{3}$ splittings in the PTO layers are smaller than those in STO layers. The value for each unit cell along the superlattice is displayed in Figure 3c. For better accuracy, the positions of the t$_{2g}$ and e$_g$ peak maxima were obtained from the first derivatives of the EELS spectra. The splitting value in the center of the PTO layers reaches 1.95~eV, significantly higher than the 1.65~eV obtained for the reference 12~nm PTO layer, and increasing further on approaching the interface.
Interestingly, a gradual broadening of the L$_3$-e$_{g}$ band is also observed when moving from the interface (spectrum 9) to the center of the PTO layer (spectrum 12).
The  spectral splitting of the STO layers is closer to the value of the STO substrate (2.3 eV), but always around $50-100$ meV smaller. In order to rule out the interdiffusion of Sr and Pb as the cause of the gradual variation in PTO, chemical profiles across the interface were obtained by multiple least squares fitting of the O-K edge signal from 536 to 540.8 eV with the two reference components (red and green spectra in Fig.3e).
The resulting Sr and Pb distributions are plotted in Fig.3b (green and red line, respectively).
At interfaces, i.e. units cells labeled 3, 9, 15 and 20, both signals clearly cross inbetween neighboring atomic columns, confining any possible Pb/Sr interdiffusion to $\pm$1 unit cell from the interface.\\
\begin{figure}
\includegraphics[width=\columnwidth]{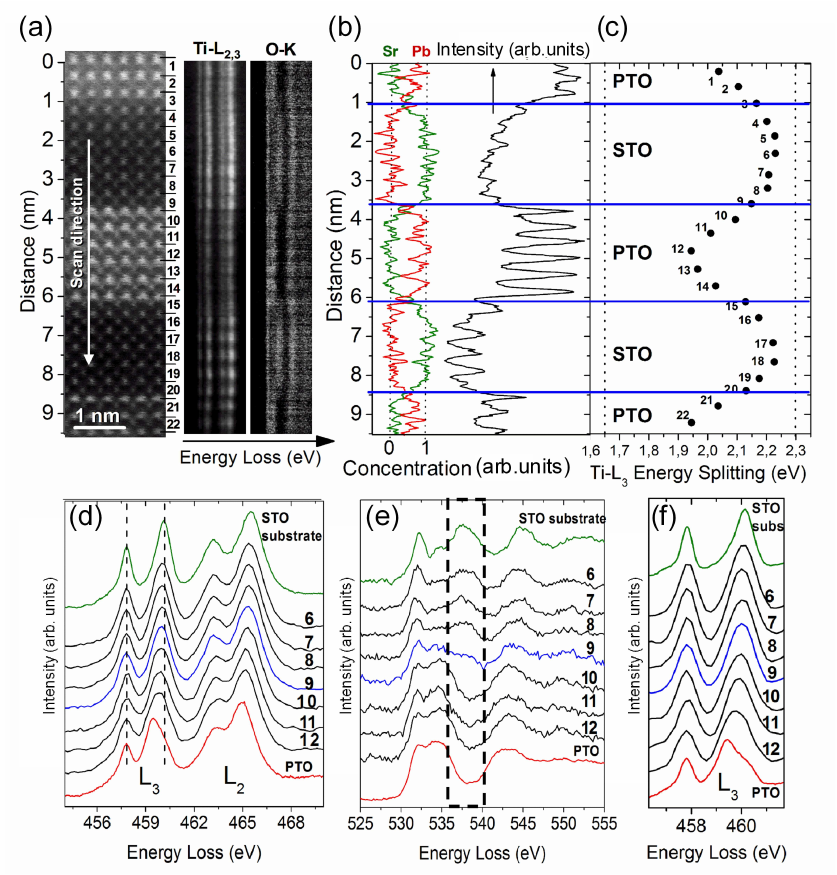}
\caption{(a) HAADF-STEM image and ELNES of Ti-L$_{2,3}$ and O-K edges across the (6$|$6) superlattice. (b) Simultaneous dark field profile recorded along the line scan (black line) and concentration profiles of Sr (green line) and Pb (red line). Horizontal lines mark the interfaces. (c) Energy splitting values in PTO and STO layers across the superlattice. Dashed lines indicate the values for STO substrate and 12~nm thick PTO. (d) High resolution Ti-L$_{2,3}$ and (e) O-K spectra across the PTO-STO interface after background substraction. (f) Ti-L$_3$ edges obtained after 10 R-L iterations.}
\end{figure}
To relate the observed continuous evolution of the STO and PTO spectra across the superlattice to structural variations, density functional theory within the local density approximation (LDA) \cite{GonzeCompPhysCommun2009} and charge transfer multliplet calculations \cite{GrootMicron2010} were performed. Two sets of structural models, one for STO and one for PTO, were built by artificially varying the $c/a$ ratio from 1 to 1.063, corresponding to the limits of bulk materials. To reproduce the experimental in-plane epitaxial constrain, we set $a=b=3.905$~\AA.

Figures 4a and 4b display the relaxed Ti and O atomic positions obtained for PTO and STO unit cells with different $c/a$ ratios, respectively. For PTO, excellent agreement was found between the experimental and relaxed atomic positions of the room temperature PTO unit cell (marked with a square) \cite{GlazerActaCrystB1978}. Note that as the $c/a$ ratio is reduced, the atomic position of Ti, O(1) and O(2) remain strongly displaced from the ideal centrosymmetric positions, except for the cubic unit cell ($c/a=1$).
Our LDA calculations for STO confirm the paraelectric state $c/a=1$ as the predicted ground state but, in this case, significant Ti off-centering is only observed once the tetragonality exceeds $\sim 1.02$. From then on, the Ti, O(1) and O(2) atoms move further away from the centrosymmetric positions, but the resulting polar distortion is much smaller than that in PTO with the same $c/a$.

The electronic density of states was computed for the whole relaxed geometry \cite{BlahaISBN}.
The projected unoccupied density of states on Ti atoms was further projected onto the Ti $d$ orbitals and decomposed according to the local symmetry C$_{4v}$, i.e., $\textit{d}_{z^2}$, \textit{d}$_{x^2-y^2}$, \textit{d}$_{xy}$ and \textit{d}$_{xz}$+\textit{d}$_{yz}$  \cite{WoickPRB2007}. This non centro-symmetric point group is induced by the ferroelectric displacement of the Ti atom along the \textit{c} axis.
The resulting values obtained for the LDA CFS (CFS = E($\textit{d}_{z^2}$, \textit{d}$_{x^2-y^2}$)-E(\textit{d}$_{xy}$, \textit{d}$_{xz}$+\textit{d}$_{yz}$)) for STO and PTO unit cells with different $c/a$ ratio are shown in Fig.4c and 4d, respectively.

These CFS together with the tetragonal distortion energy differences ($\Delta_1$=E(\textit{d}$_{xy}$)-E(\textit{d}$_{xz}$+\textit{d}$_{yz}$) and $\Delta_2$=E($\textit{d}_{z^2}$)-E(\textit{d}$_{x^2-y^2}$)), displayed in figures 4c and 4d, and the estimated band widths obtained by LDA for the STO and PTO were used as input data for charge transfer multiplet calculations, yielding a qualitatively good match with the experimental measurements (Fig.4e). In particular, the difference between the energy splittings of PTO and STO ($\Delta$E(L$_3$)$_{STO}$-$\Delta$E(L$_3$)$_{PTO}$) is computed to be 0.60~eV, close to the measured value 0.65~eV.
Moreover, the asymmetric broadening in the L$_3$-e$_{g}$ band of the PTO simulated-spectra with respect to STO spectra agrees with that experimentally observed in the PTO layers (spectra 9 to 12 in Fig.3f).

We have also performed similar calculations for PTO and STO structures with a tetragonal distortion but forcing a paraelectric state (atomic displacements $\delta=0$), by using the Ti centrosymmetric local symmetry D$_{4h}$. The LDA calculations result in a smaller cubic crystal field evolution but much larger tetragonal distortion energy that should induce some spectral difference in the EELS data. Although the charge transfer multiplet calculations of the Ti L$_{2,3}$ edges using ab-initio computed parameters are not reliable enough for a quantitative match with experiment, it can provide valuable information about the spectral differences expected between FE and PE state. Figure 4f shows the Ti-L$_{2,3}$ spectra as obtained by the charge transfer multiplet calculations for PE and FE tetragonally distorted PTO unit cell ($c/a$=1.063). The main splitting of the L$_3$ and of the L$_2$ line do not vary much, but features appear due to the strong D$_{4h}$ distortion parameters. Such features are not observed neither in our experimental EELS data nor in previous reported XAS data of tetragonaly distorted STO or PTO \cite{WoickPRB2007,Arenholz2010}. These results thus exclude a possible centrosymmetric arrangement of the atoms in the PTO layers, confirming their ferroelectric nature, as already revealed by the presence of ferroelectric domains. The large cell-to-cell variation of the CFS in the PTO layers can thus be attributed to a variation in tetragonality and polarization, showing that the uniform polarization model, frequently used to describe monodomain superlattices \cite{NeatonRabe2003} does not apply to our polydomain samples.

Combining the results of the \textit{ab-initio} calculations and the measurements of the reference PTO and STO spectra allows us to estimate the local distortions within the superlattice layers. The CFS of the central PTO layer in the superlattice is 0.3~eV larger than that of the reference PTO layer, corresponding to a significantly smaller $c/a$ value of about 1.025 which further decreases towards the interface.
The STO layers show a weaker CFS variation, with the four central unit cells all showing splittings around 0.1~eV below that of the cubic STO substrate. The small tetragonality is consistent with X-ray diffraction measurements on pure paraelectric STO films, which yield a c-axis lattice parameter of 3.92~\AA\, possibly signalling a slight off-stoichiometry \cite{BrooksAPL2009}. The very weak distortion in the STO makes it difficult to distinguish whether these layers are polar or not but if any polarization is present, it is small and any electrostatic coupling between the PTO layers \cite{StephanovichPRL2005} is therefore likely to be weak.
The structural distortions illustrated in Fig.3c may arise from the inhomogenous strains associated with the domain structure itself. However, we should note that a similar reduction of tetragonality at interfaces or surfaces of Pb-based ferroelectrics has also been observed in thin films without stripe domains \cite{JiaNatMater2007} and hence the precise physical origin of these distortions requires further study.\\
\begin{figure}
\includegraphics[width=\columnwidth]{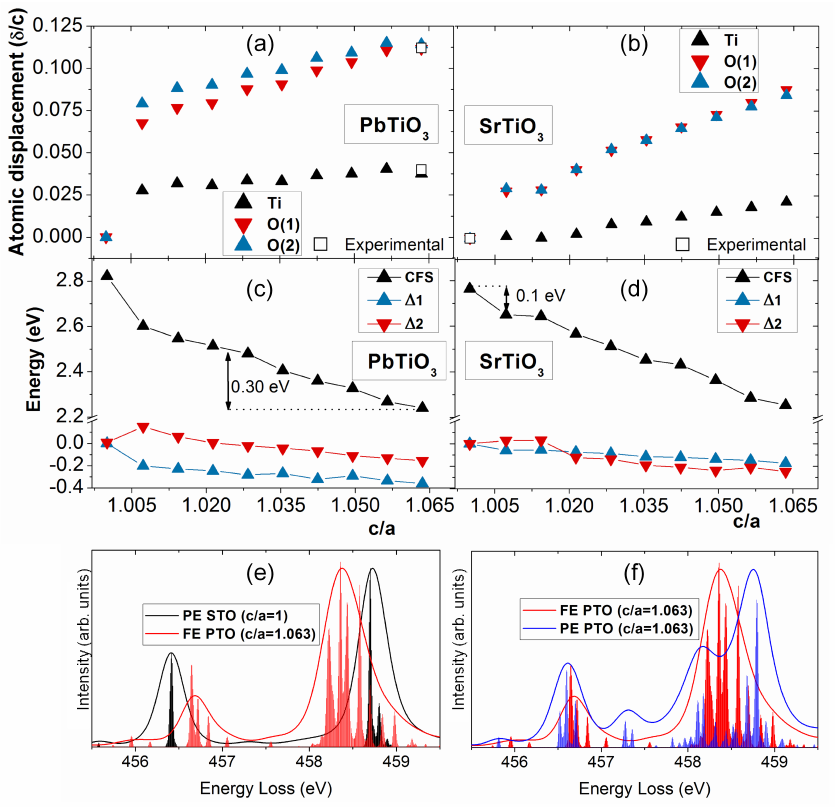}
\caption{Relaxed atomic displacements as a function of the $c/a$ ratio for (a) PTO and (b) STO structural models. CFS values for (c) PTO and (d) STO structural models as a function of the $c/a$ ratio. (e) Simulated L$_3$ spectra for PTO (red) and STO (black). (f) Simulated L$_3$ spectra for a paraelectric (blue) and polar (red) tetragonally distorted PTO unit cell ($c/a$=1.063).}
\end{figure}
In this article, we have illustrated that high resolution EELS is a powerful alternative technique for studying local ferroelectric distortions at the perovskite unit-cell scale. The technique is sensitive enough to measure small variations in the L$_{2,3}$-edge down to 50 meV that can be semi-quantitatively related to a $c/a$ evolution as small as $1\%$, allowing us to map with ultra high sensitivity cell-to-cell variations in local structural distortions and revealing an inhomogeneous polarization profile within the ferroelectric and paraelectric layers that previously has only been predicted theoretically or inferred by indirect means.
Recent XAS measurements reveal that linear dichroism is sensitive to the polarization direction \cite{Arenholz2010}, suggesting that this information could also be accessible through anisotropic EELS. However, achieving high spatial resolution in anisotropic EELS measurements is still a challenging task.\\
The authors acknowledge funding from the Swiss National Science Foundation through the NCCR MaNEP and division II, the EU OxIDes project and the ESTEEM program.

\end{document}